\documentclass[aps,prd,preprint,showpacs,nofootinbib,11pt]{revtex4-1}
\usepackage{latexsym,bm,amsmath,amssymb}

\begin{document}

\title{Conserved Killing charges of quadratic curvature gravity theories in arbitrary backgrounds} 

\author{Deniz Olgu Devecio\u{g}lu}
\email{e137112@metu.edu.tr}
\author{{\"O}zg{\"u}r Sar{\i}o\u{g}lu}
\email{sarioglu@metu.edu.tr}
\affiliation{Department of Physics, Faculty of Arts and  Sciences,\\
             Middle East Technical University, 06531, Ankara, Turkey}

\date{\today}

\begin{abstract}
We extend the Abbott-Deser-Tekin procedure of defining conserved quantities of asymptotically constant-curvature spacetimes, 
and give an analogous expression for the conserved charges of geometries that are solutions of quadratic curvature gravity 
models in generic $D$-dimensions and that have arbitrary asymptotes possessing at least one Killing isometry. We show 
that the resulting charge expression correctly reduces to its counterpart when the background is taken to be a space of constant 
curvature and, moreover, is background gauge invariant. As applications, we compute and comment on the energies of two 
specific examples: the three dimensional Lifshitz black hole and a five dimensional companion of the first, whose energy has 
never been calculated beforehand.
\end{abstract}


\maketitle

There has been an ongoing interest in gravitational models that involve quadratic curvature terms which naturally emerge
in various string theories or quantum gravity models. A recent example of these with physically interesting properties is
the three dimensional New Massive Gravity (NMG) of \cite{hohm}. There are also a growing number of exact analytic 
solutions to these models which asymptote to rather exotic geometries other than the more familiar spaces of constant 
curvature. An interesting family of such geometries involve the so called Lifshitz black holes in various dimensions 
\cite{giri1,giri2} of late, which were inspired by the four-dimensional Lifshitz black hole solution of \cite{cinli}.
These spacetimes have emerged as gravity duals of some nonrelativistic scale invariant condensed matter theories via a 
generalization of the AdS/CFT correspondence to such systems (see \cite{giri1,giri2} and the references therein for 
details). It is obviously of importance to have a better understanding of the physical properties of these solutions. 

The main aim of the present work is to provide a step further in this direction. For this purpose we generalize the 
Abbott-Deser-Tekin (ADT) procedure \cite{des1,des2,des3} originally developed for working with spacetimes that 
asymptote to spaces of constant curvature to spacetimes that have arbitrary asymptotic behavior and that are solutions 
of quadratic curvature gravity theories in generic $D$-dimensions. We show how the general formula reduces to the 
original ADT expression of \cite{des3} in the limit when the background is a space of constant curvature. To illustrate 
the idea, we apply this new charge definition to compute the energy of the three dimensional Lifshitz black hole \cite{giri1}, 
which was calculated by other methods beforehand \cite{toni} and comment on the result. As another application, we 
also find the energy of a five dimensional Lifshitz black hole \cite{giri2}, which has not been computed before.

Let us first review the original ADT approach for finding the conserved gravitational charges of a generic gravity model. 
One should start with a local gravity action whose field equations read
\begin{equation}
 \Phi_{ab} = \kappa \, \tau_{ab} \, , \label{feqns}
\end{equation}
where $\tau_{ab}$ describes the energy-momentum of a covariantly conserved source and $\kappa$ is basically a
gravitational coupling constant. Now suppose that there exists a background metric $\bar{g}_{ab}$ which satisfies
\( \bar{\Phi}_{ab}(\bar{g}) = 0 \) for \( \tau_{ab}= 0 \). One can linearize a generic metric $g_{ab}$ which asymptotically
approaches to the background $\bar{g}_{ab}$ in the usual fashion as
\begin{equation}
 g_{ab} = \bar{g}_{ab} +  h_{ab} \, .
\end{equation}
Here the deviation $h_{ab}$ should vanish sufficiently rapidly as one approaches the background at infinity. Linearizing 
the full field equations (\ref{feqns}) about the background, one obtains
\begin{equation}
 \Phi_{ab}^{L} = \kappa \, T_{ab} \, , \label{fleqns}
\end{equation}
where $\Phi_{ab}^{L}$ is linear in $h_{ab}$ and the right hand side of (\ref{fleqns}) contains all the remaining nonlinear
parts as well as $\tau_{ab}$. One should keep in mind that the background metric $\bar{g}_{ab}$ is responsible for
raising and lowering indices and is used in defining the covariant derivative $\bar{\nabla}_{a}$. Let us now assume that
the background metric $\bar{g}_{ab}$ admits at least one globally defined\footnote{Here by \textquotedblleft globally 
defined\textquotedblright, we mean that the Killing vector is smooth and well-defined in the entire local coordinate chart
that the metric $\bar{g}_{ab}$ works.} Killing vector $\bar{\xi}^{a}$ such that
\begin{equation}
 \bar{\nabla}_{a} \bar{\xi}_{b} +  \bar{\nabla}_{b} \bar{\xi}_{a} = 0 \, . \label{kill}
\end{equation}

As with the original ADT approach \cite{des1,des2,des3}, one now needs the condition \( \bar{\nabla}_{a} T^{ab} = 0 \),
which follows from the Bianchi identity of the full theory, just as in the case of a constant curvature background. Now
using this with the defining property of a Killing vector (\ref{kill}), one arrives at the usual expression
\begin{equation}
 \bar{\nabla}_{a} (T^{ab} \, \bar{\xi}_{b}) = 0 = \partial_{a} (\sqrt{-\bar{g}} \, T^{ab} \, \bar{\xi}_{b}) \,, \label{cons}
\end{equation}
which one can use to construct a conserved Killing charge as in \cite{des1,des2,des3} provided $T^{ab} \, \bar{\xi}_{b}$
can be cast as a divergence, i.e. \( T^{ab} \, \bar{\xi}_{b} = \bar{\nabla}_{b} {\mathcal F}^{ab} \) for an antisymmetric
tensor ${\mathcal F}^{ab}$ modulo terms that vanish on-shell. In that case the required charge expression takes the form
\begin{equation}
 Q^{a}(\bar{\xi}) = \int_{\Sigma} d^{D-1}x \, \sqrt{-\bar{g}} \, \Phi^{ab}_{L} \, \bar{\xi}_{b} =
 \int_{\partial \Sigma} d \bar{S}_{b} {\mathcal F}^{ab} \,, \label{genchar}
\end{equation}
where $\Sigma$ denotes the ($D-1$)-dimensional hypersurface of the $D$-dimensional spacetime and $\partial \Sigma$
is its ($D-2$)-dimensional boundary with the surface element $d \bar{S}$ \footnote{Here we refrain from delving into 
delicate issues involving mathematical rigor and assume implicitly the validity of all steps taken in arriving at (\ref{genchar}).}.

Let us now see how one can define conserved Killing charges (\ref{genchar}) of a generic quadratic curvature gravity 
theory described by the action\footnote{Throughout we use conventions in which the signature of metrics is \( (-,+,+,\dots) \), 
all covariant derivatives satisfy \( [\nabla_{a}, \nabla_{b}] V_{c} = R_{abc}\,^{d} \, V_{d} \) and the Ricci tensor is 
related to the Riemann tensor as \( R_{ab} = R^{c}\,_{acb} \). We also take the symmetric and the antisymmetric parts 
of any 2-index tensor $W_{ab}$ as \( W_{(ab)} \equiv \frac{1}{2} (W_{ab} + W_{ba}) \) and 
\( W_{[ab]} \equiv  \frac{1}{2} (W_{ab} - W_{ba}) \), respectively.}
\begin{equation}
 I = \int \, d^{D}x \, \sqrt{-g} \, \Big( \frac{1}{\kappa} (R + 2 \Lambda_{0}) + \alpha R^{2} 
+ \beta R_{ab} R^{ab} + \gamma (R_{abcd} R^{abcd} - 4 R_{ab} R^{ab} + R^2) \Big), \label{act}
\end{equation}
where $\Lambda_{0}$ is the bare cosmological constant and $\kappa$ is related to the $D$-dimensional
Newton's constant $G_{D}$. The field equations following from the variation of the action $I$ yield
\begin{equation}
 \Phi_{ab} \equiv \frac{1}{\kappa} {\mathcal G}_{ab} + \alpha A_{ab} + \beta B_{ab} + \gamma C_{ab} \,, \label{feq}
\end{equation}
where
\begin{eqnarray}
 {\mathcal G}_{ab} & \equiv & R_{ab} - \frac{1}{2} g_{ab} R - \Lambda_{0} g_{ab}, \\
 A_{ab} & \equiv & 2 R R_{ab} - 2 \nabla_{a} \nabla_{b} R + g_{ab} (2 \square R - \frac{1}{2} R^2), \\
 B_{ab} & \equiv & 2 R_{acbd} R^{cd} - \nabla_{a} \nabla_{b} R +  \square R_{ab} + \frac{1}{2} g_{ab} (\square R - R_{cd} R^{cd}) , \\ 
 C_{ab} & \equiv & 2 R R_{ab} - 4 R_{acbd} R^{cd} + 2 R_{acde} R_{b}\,^{cde} - 4 R_{ac} R_{b}\,^{c}
 - \frac{1}{2} g_{ab} (R_{cdef} R^{cdef} - 4 R_{cd} R^{cd} + R^{2}) , \label{eqns}
\end{eqnarray}
where \( \square \equiv \nabla_{c} \nabla^{c} \). For $D<5$, $C_{ab}$, coming from the 
Gauss-Bonnet term, automatically vanishes.

As discussed before, the field equations (\ref{feq})-(\ref{eqns}) need to be linearized about a
general background to first order in $h_{ab}$. One simply finds
\begin{equation}
 \Phi_{ab}^{L} \equiv \frac{1}{\kappa} {\mathcal G}_{ab}^{L} + \alpha A_{ab}^{L} 
 + \beta B_{ab}^{L} + \gamma C_{ab}^{L} \,, \label{fleq}
\end{equation}
analogous to (\ref{feq}), where now
\begin{eqnarray}
 {\mathcal G}_{ab}^{L} & \equiv & R_{ab}^{L} - \frac{1}{2} \bar{g}_{ab} R_{L} - \frac{1}{2} h_{ab} \bar{R} 
- \Lambda_{0} h_{ab} , \label{lein} \\
 A_{ab}^{L} & \equiv & 2 (\bar{R} R_{ab}^{L} + \bar{R}_{ab} R_{L}) - 2 (\nabla_{b} \nabla_{a} R)_{L} 
 + \bar{g}_{ab} (2 (\square R)_{L} -  \bar{R} R_{L}) + h_{ab} (2 \bar{\square} \bar{R} - \frac{1}{2} \bar{R}^2) , \\
 B_{ab}^{L} & \equiv & 2 (R_{acbd} R^{cd})_{L} - (\nabla_{b} \nabla_{a} R)_{L} + (\square R_{ab})_{L} 
 + \frac{1}{2} \bar{g}_{ab} ((\square R)_{L} - (R_{cd} R^{cd})_{L}) \nonumber \\
 & & + \frac{1}{2} h_{ab} (\bar{\square} \bar{R} - \bar{R}_{cd} \bar{R}^{cd}) , \\
 C_{ab}^{L} & \equiv &  2 (\bar{R} R_{ab}^{L} + \bar{R}_{ab} R_{L}) - 4 (R_{acbd} R^{cd})_{L} 
 + 2 (R_{acde} R_{b}\,^{cde})_{L} - 4 (R_{ac} R_{b}\,^{c})_{L} \nonumber \\ 
 & & - \frac{1}{2} \bar{g}_{ab} \big( (R_{cdef} R^{cdef})_{L}  - 4 (R_{cd} R^{cd})_{L} + 2 \bar{R} R_{L} \big) 
 - \frac{1}{2} h_{ab} (\bar{R}_{cdef} \bar{R}^{cdef} - 4 \bar{R}_{cd} \bar{R}^{cd} + \bar{R}^2) . \label{leqns}
\end{eqnarray}
At this stage we refer the reader to appendix \ref{appa}, where it is shown how the ingredients of equations (\ref{lein})-(\ref{leqns})
can be calculated recursively using the Cadabra software \cite{cad1,cad2}, to obtain the final form of the linearized 
field equations $\Phi^{ab}_{L}$.

Continuing in the fashion described beforehand, one finds that
\begin{equation}
 \Phi^{ab}_{L} \bar{\xi}_{b} = \bar{\nabla}_{b} {\mathcal F}^{ab} + h^{ab} \bar{\Phi}_{bc} \bar{\xi}^{c} 
+  \frac{1}{2} \bar{\xi}^{a} \bar{\Phi}_{bc} h^{bc} - \frac{1}{2} h \bar{\Phi}^{ab} \bar{\xi}_{b} \,, \label{phixi}
\end{equation}
where
\begin{equation}
 {\mathcal F}^{ab} = \frac{1}{\kappa} {\mathcal F}^{ab}_{E} + \alpha {\mathcal F}^{ab}_{\alpha} 
 + \beta {\mathcal F}^{ab}_{\beta} + \gamma {\mathcal F}^{ab}_{\gamma} \,, \label{fab}
\end{equation}
with
\begin{eqnarray}
 {\mathcal F}^{ab}_{E} & \equiv & \bar{\xi}_{c} \bar{\nabla}^{[a} h^{b]c} + \bar{\xi}^{[b} \bar{\nabla}_{c} h^{a]c} 
 + h^{c[b} \bar{\nabla}_{c} \bar{\xi}^{a]} + \bar{\xi}^{[a} \bar{\nabla}^{b]} h + \frac{1}{2} h \bar{\nabla}^{[a} \bar{\xi}^{b]} , 
 \label{einchar} \\
 {\mathcal F}^{ab}_{\alpha} & \equiv & 2 \bar{R} {\mathcal F}^{ab}_{E} + 2 \bar{\xi}^{[b} h^{a]c} \bar{\nabla}_{c} \bar{R}
 + 4 \bar{\xi}^{[a} \bar{\nabla}^{b]} R_{L} + 2 R_{L} \bar{\nabla}^{[a} \bar{\xi}^{b]} , \label{alchar} \\
 {\mathcal F}^{ab}_{\beta} & \equiv & \bar{\xi}^{[a} \bar{\nabla}^{b]} R_{L} + 2 \bar{\xi}^{c} \bar{\nabla}^{[b} (R^{a]}\,_{c})_{L} 
 + 2 (R^{[b}\,_{c})_{L} \bar{\nabla}^{a]} \bar{\xi}^{c} + h_{cd} \bar{\xi}^{[b} \bar{\nabla}^{a]} \bar{R}^{cd} 
 + 2 h^{c[a} \bar{\xi}_{d} \bar{\nabla}_{c} \bar{R}^{b]d} + 2 \bar{R}^{c[a} \bar{\xi}_{d} \bar{\nabla}_{c} {h^{b]d}} \nonumber \\
 & & + h \bar{\xi}_{c} \bar{\nabla}^{[b} \bar{R}^{a]c} + 2 \bar{R}^{c[b} h^{a]d} \bar{\nabla}_{c} \bar{\xi}_{d} 
 + h \bar{R}^{c[b} \bar{\nabla}^{a]} \bar{\xi}_{c} + 2 \bar{R}^{c[b} \bar{\xi}^{d} \bar{\nabla}_{d} {h^{a]}\,_{c}} 
 + \bar{\xi}^{[a} \bar{R}^{b]c} \bar{\nabla}_{c} h + \bar{R}^{cd} \bar{\xi}^{[a} \bar{\nabla}^{b]} h_{cd} \nonumber \\
 & & + 2 \bar{\xi}^{d} \bar{R}_{cd} \bar{\nabla}^{[b} h^{a]c} + 2 \bar{R}^{c[a} \bar{\xi}^{b]} \bar{\nabla}^{d} h_{cd} 
 + 2 \bar{\xi}^{d} \bar{R}^{c[b} \bar{\nabla}^{a]} h_{cd} , \label{bechar} \\
 {\mathcal F}^{ab}_{\gamma} & \equiv & 2 \bar{R} {\mathcal F}^{ab}_{E} + 2 \bar{R}^{[ba]cd} \bar{\xi}_{d} \bar{\nabla}_{c} h
 + 4 \bar{\xi}_{c} \bar{R}^{c[b} \bar{\nabla}^{a]} h + 4 \bar{R}^{c[a} \bar{\xi}^{b]} \bar{\nabla}_{c} h 
 + 2 h \bar{R}^{c[ab]d} \bar{\nabla}_{d} \bar{\xi}_{c} + 4 h \bar{R}^{c[a} \bar{\nabla}^{b]} \bar{\xi}_{c} \nonumber \\
 & & + 4 \bar{\xi}_{d} \bar{R}^{dec[a} \bar{\nabla}_{c} h^{b]}\,_{e} + 4 \bar{\xi}_{d} \bar{R}^{dec[a} \bar{\nabla}_{e} h^{b]}\,_{c} 
 + 4 \bar{\xi}_{d} \bar{R}^{dec[b} \bar{\nabla}^{a]} h_{ce} + 4 \bar{\xi}^{[a} \bar{R}^{b]cde} \bar{\nabla}_{d} h_{ce} 
 + 2 \bar{R}^{[ab]cd} \bar{\xi}^{e} \bar{\nabla}_{c} h_{de} \nonumber \\
 & & + 2 \bar{\xi}_{d} h_{ce} \bar{\nabla}^{c} \bar{R}^{[ab]de} + 4 \bar{\xi}_{d} \bar{R}^{d[a} \bar{\nabla}_{c} h^{b]c} 
 + 4 h_{ce} \bar{R}^{dec[a} \bar{\nabla}^{b]} \bar{\xi}_{d} + 4 \bar{R}^{[ab]cd} h_{ce} \bar{\nabla}_{d} \bar{\xi}^{e} 
 + 4 \bar{R}_{cd} \bar{\xi}^{d} \bar{\nabla}^{[b} h^{a]c} \nonumber \\
 & & + 4 \bar{\xi}^{d} \bar{R}^{c[a} \bar{\nabla}^{b]} h_{cd} + 4 \bar{\xi}^{d} \bar{R}_{c}\,^{[b} \bar{\nabla}_{d} h^{a]c}
 + 4 \bar{\xi}^{d} \bar{R}^{c[b} \bar{\nabla}_{c} h^{a]}\,_{d} + 8 \bar{R}^{cd} \bar{\xi}^{[a} \bar{\nabla}_{c} h^{b]}\,_{d} 
 + 4 \bar{R}^{cd} \bar{\xi}^{[b} \bar{\nabla}^{a]} h_{cd} \nonumber \\
 & & + 2 \bar{R}^{cd} h_{cd} \bar{\nabla}^{[b} \bar{\xi}^{a]} + 4 \bar{\xi}_{d} h^{c[a} \bar{\nabla}_{c} \bar{R}^{b]d} 
 + 4 h^{cd} \bar{\xi}^{[b} \bar{\nabla}_{c} \bar{R}^{a]}\,_{d} + 4 h^{c[a} \bar{R}^{b]d} \bar{\nabla}_{c} \bar{\xi}_{d} 
 + 8 h_{cd} \bar{R}^{c[b} \bar{\nabla}^{a]} \bar{\xi}^{d} \nonumber \\
 & & + 2 \bar{\xi}^{[a} h^{b]c} \bar{\nabla}_{c} \bar{R} , \label{gachar}
\end{eqnarray}
and one should recall that the background was chosen such that \( \bar{\Phi}_{ab}(\bar{g}) = 0 \) so that (\ref{genchar}) follows.

There are a few nontrivial requirements that the gravitational charge $Q^{a}(\bar{\xi})$  given by (\ref{genchar})
(with (\ref{fab})-(\ref{gachar})) must satisfy. The first one is that it has to reduce to its counterpart given in
\cite{des2,des3} when the background $\bar{g}_{ab}$ is taken to be a space of constant curvature. In that case 
one simply sets
\[ \bar{R}_{abcd} = \frac{2 \Lambda}{(D-2)(D-1)} (\bar{g}_{ac} \bar{g}_{bd} - \bar{g}_{ad} \bar{g}_{bc}) , \quad
 \bar{R}_{ab} = \frac{2 \Lambda}{D-2} \bar{g}_{ab} , \quad \bar{R} =  \frac{2 D \Lambda}{D-2} , \]
where $\Lambda$ is the effective cosmological constant (see \cite{des3} for details), in (\ref{einchar}) to (\ref{gachar}). 
We have verified explicitly that this is indeed the case and $Q^{a}(\bar{\xi})$ reduces to equation (31) of \cite{des3} as
it should. The second requirement is that $Q^{a}(\bar{\xi})$ has to be background gauge invariant and this is
shown in appendix \ref{appb} below.

We should also make a few remarks regarding the tensor potentials (\ref{einchar}) to (\ref{gachar}). Equations (\ref{phixi}) 
to (\ref{einchar}), in the case \( \alpha = \beta = \gamma = 0 \), have already appeared in \cite{cle} in the context of defining
the conserved Killing charges of Topologically Massive Gravity (TMG) \cite{des4} around an arbitrary background. Similarly
equations (\ref{phixi}) to (\ref{alchar}), in the case \( \beta = \gamma = 0 \), have been presented in \cite{kore} where 
an analogous task was undertaken for the NMG theory of \cite{hohm}. In \cite{kore}, the authors have also
given an expression (see equation (27) of \cite{kore}) in the case \( \gamma = 0 \) which should correspond to our tensor potential
${\mathcal F}^{ab}_{\beta}$ (\ref{bechar}). However the expression they have is considerably different than ours. We strongly 
believe that our ${\mathcal F}^{ab}_{\beta}$ is correct since it already passes the two nontrivial requirements mentioned above. 
Further work is required to understand the discrepancy between the two results. As a final remark, the tensor potential 
${\mathcal F}^{ab}_{\gamma}$ (\ref{gachar}), to our knowledge, has not appeared elsewhere beforehand.

Let us now consider a few illustrative examples where the charge definition $Q^{a}(\bar{\xi})$ (\ref{genchar}) can be
employed to demonstrate its practical use. For this purpose we resort to the Lifshitz black holes that we have alluded
to earlier.

The three dimensional Lifshitz black hole of \cite{giri1}, which reads
\begin{equation}
 ds^2 = - \frac{r^6}{\ell^6} \Big(1 - \frac{m \ell^2}{r^2} \Big) dt^2 + \frac{dr^2}{\frac{r^2}{\ell^2} - m} + \frac{r^2}{\ell^2} dx^2, 
 \label{lif3d}
\end{equation}
is a solution to the cosmological NMG theory of \cite{hohm} with
\[ \Lambda_{0} = \frac{13}{2 \ell^2}, \quad \beta = \frac{2 \ell^2}{\kappa}, \quad \alpha = - \frac{3 \ell^2}{4 \kappa} \]
in our conventions. Taking the background to be the Lifshitz spacetime
\[ ds^2 = - \frac{r^6}{\ell^6} dt^2 + \frac{\ell^2}{r^2} dr^2 + \frac{r^2}{\ell^2} dx^2, \]
which is obtained by the $m \to 0$ limit in (\ref{lif3d}), employing the timelike Killing vector \( \bar{\xi}^{a} = (-1,0,0) \),
setting $\kappa = 16 \pi G$ in accordance with the conventions of \cite{giri1} and noting that the angular coordinate $x$ is
periodic with $2 \pi \ell$, one finds the energy of (\ref{lif3d}) to be
\begin{equation}
 E = \lim_{r \to \infty} \int_{0}^{2 \pi \ell} \frac{m^2 r^2 (- 9 \ell^4 m^2 + 24 \ell^2 m r^2 - 7 r^4)}{\ell \kappa (\ell^2 m -r^2)^3} dx
 = \int_{0}^{2 \pi \ell} \frac{7 m^2}{\ell \kappa} dx = \frac{7 m^2}{8 G}, \label{enlif}
\end{equation}
where $G$ denotes the three dimensional Newton's constant.

The energy of (\ref{lif3d}) has also been computed in \cite{toni} by using the boundary stress tensor method, and found to
be \(E = m^2/4 G \). Clearly there is a discrepancy between this and our result, which deserves further attention.

Another nontrivial example involving a five dimensional Lifshitz black hole can also be given. For this purpose we
consider the spacetime given in section 3.1 (in the case $z=2$) of \cite{giri2} which reads
\begin{equation}
 ds^2 = - \frac{r^4}{\ell^4} \Big(1 - \frac{m \ell^{5/2}}{r^{5/2}} \Big) dt^2 
 + \frac{\ell^2}{r^2} \Big(1 - \frac{m \ell^{5/2}}{r^{5/2}} \Big)^{-1} dr^2 
 + \frac{r^2}{\ell^2} d\vec{x}^2, 
 \label{5dlif}
\end{equation}
with
\[ \kappa = 1, \quad \Lambda_{0} = \frac{2197}{551 \ell^2}, \quad \alpha = - \frac{16 \ell^2}{725}, 
   \quad \beta = \frac{1584 \ell^2}{13775}, \quad \gamma = \frac{2211 \ell^2}{11020} \]
in our conventions. As in the previous example we take the background to be given by the Lifshitz spacetime obtained by
taking the $m \to 0$ limit in (\ref{5dlif}) and the timelike Killing vector to be \( \bar{\xi}^{a} = (-1,0,0,0,0) \). With these
choices the energy of (\ref{5dlif}) turns out to be 
\begin{equation}
 E = \frac{536 m^2 \Omega}{2755 \ell}, \label{en5dlif}
\end{equation}
where $\Omega$ denotes the contribution of the three dimensional integration over the ranges of the angular variables $\vec{x}$.
To our knowledge this is the first time that the energy of (\ref{5dlif}) has ever been calculated.

As a summary, we have generalized the ADT charge definition to work with spacetimes that have arbitrary asymptotic 
behavior for the case of quadratic curvature gravity theories in generic $D$-dimensions. We have checked that this 
definition reduces to the original ADT expression when the background is a space of constant curvature. We have 
also shown that it is background gauge invariant. We have seen how it works on two illustrative examples.

Clearly there are various open problems that can be attacked using our work. It would be interesting to compute 
the charges of other exotic solutions of NMG that have not been considered here using our $Q^{a}(\bar{\xi})$ 
(\ref{genchar}). Here we have considered only one example from the solutions listed in \cite{giri2}. The others
certainly deserve further attention. It may be of interest to apply the ADT procedure generalized for generic
quadratic curvature gravity models here to the so called Lovelock gravity theories of late.

\begin{acknowledgments}
We would like to thank Bayram Tekin for useful discussions and his careful reading of this manuscript. We would also
like to thank the anonymous referee whose useful comments led us to correct a mistake in an earlier version of this
paper. This work is partially supported by the Scientific and Technological Research Council of Turkey (T{\"U}B\.{I}TAK).
\end{acknowledgments}

\appendix

\section{\label{appa} The linearized field equations $\Phi^{ab}_{L}$}
Here we list some useful identities that have been used in the linearization of the full field equations
(\ref{feq})-(\ref{eqns}) and describe how one can obtain $\Phi^{ab}_{L}$.

The inverse of the metric is given by \( g^{ab} = \bar{g}^{ab} -  h^{ab} \) to ${\mathcal O}(h^{2})$. Using this,
one can find the linearized Christoffel symbols as
\begin{equation}
 (\Gamma^{a}\,_{bc})_{L} = \frac{1}{2} \bar{g}^{ad} 
 \left( \bar{\nabla}_{b} h_{cd} + \bar{\nabla}_{c} h_{bd} - \bar{\nabla}_{d} h_{bc} \right) . \label{chilin}
\end{equation}
A straightforward calculation shows that the linearized Riemann tensor is given by
\begin{equation}
 (R^{a}\,_{bcd})_{L} =  \bar{\nabla}_{c} (\Gamma^{a}\,_{bd})_{L} -  \bar{\nabla}_{d} (\Gamma^{a}\,_{bc})_{L} ,
\label{rielin}
\end{equation}
which leads to
\begin{equation}
 R_{ab}^{L} =  \frac{1}{2} \left( \bar{\nabla}^{c} \bar{\nabla}_{b} h_{ac} 
 + \bar{\nabla}^{c} \bar{\nabla}_{a} h_{bc} - \bar{\square} h_{ab}
 - \bar{\nabla}_{a} \bar{\nabla}_{b} h \right) \label{riclin}
\end{equation}
for the linearized Ricci tensor, where \( h \equiv \bar{g}^{cd} h_{cd} \) and 
\( \bar{\square} \equiv \bar{\nabla}_{c} \bar{\nabla}^{c} \). Since the curvature scalar is defined as
\( R \equiv g^{ab} R_{ab} \), its linearized version follows as
\begin{equation}
 R_{L} = \bar{\nabla}^{a} \bar{\nabla}^{b} h_{ab} - \bar{\square} h - h^{ab} \bar{R}_{ab} . \label{ricscalin}
\end{equation}

However the covariant derivatives and the contractions of the curvature tensors also show up in the full field equations
(\ref{feq})-(\ref{eqns}), and these are to be linearized accordingly. These can be calculated in a similar fashion. We
hereby list some of the most relevant expressions involving the linearization of such terms:
\begin{eqnarray}
 (\nabla_{a} R_{cd})_{L} & = & \bar{\nabla}_{a} (R_{cd})_{L} - (\Gamma^{e}\,_{ac})_{L} \bar{R}_{ed}
 - (\Gamma^{e}\,_{ad})_{L} \bar{R}_{ec}, \nonumber \\
 (\nabla_{b} \nabla_{a} R_{cd})_{L} & = & \bar{\nabla}_{b} (\nabla_{a} R_{cd})_{L} 
 -  \bar{\nabla}_{e} \bar{R}_{cd} (\Gamma^{e}\,_{ba})_{L} 
 - \bar{\nabla}_{a} \bar{R}_{ed} (\Gamma^{e}\,_{bc})_{L} - \bar{\nabla}_{a} \bar{R}_{ce} (\Gamma^{e}\,_{bd})_{L}, \nonumber \\
 (\square R_{cd})_{L} & = & \bar{g}^{ab} (\nabla_{b} \nabla_{a} R_{cd})_{L}
 - h^{ab} \bar{\nabla}_{b} \bar{\nabla}_{a} \bar{R}_{cd}, \nonumber \\
 (\nabla_{b} \nabla_{a} R)_{L} & = & \bar{g}^{cd} (\nabla_{b} \nabla_{a} R_{cd})_{L}
 - h^{cd} \bar{\nabla}_{b} \bar{\nabla}_{a} \bar{R}_{cd}, \label{var} \\ 
 (\square R)_{L} & = & \bar{g}^{ab} (\nabla_{b} \nabla_{a} R)_{L} 
 - h^{ab} (\bar{\nabla}_{b} \bar{\nabla}_{a} \bar{R}), \nonumber \\
 (R_{acbd} R^{cd})_{L} & = & \bar{g}_{ae} (R^{e}\,_{cbd})_{L} \bar{R}^{cd} + h_{ae} \bar{R}^{e}\,_{cbd} \bar{R}^{cd} 
 + \bar{R}_{a}\,^{c}\,_{b}\,^{d} (R_{cd})_{L} - h^{ce} \bar{R}_{acbd} \bar{R}_{e}\,^{d} 
 - h^{de} \bar{R}_{acbd} \bar{R}^{c}\,_{e}, \nonumber \\
 (R_{ab} R^{ab})_{L} & = & 2 \bar{R}^{ab} (R_{ab})_{L} - 2 h^{ab} \bar{R}_{ac} \bar{R}_{b}\,^{c}. \nonumber
\end{eqnarray}
As explained in the text, we have used these and a number of such expressions recursively in the Cadabra software 
\cite{cad1,cad2}, which is rather handy for such symbolic manipulations, and obtained the final form of the linearized 
field equations $\Phi^{ab}_{L}$. However, its final form is rather cumbersome and hardly illuminating, so it is best not 
displayed here!

\section{\label{appb} The background gauge invariance of $Q^{a}(\bar{\xi})$}
Here we show how the gravitational charge $Q^{a}(\bar{\xi})$ given by (\ref{genchar}) (with (\ref{fab})-(\ref{gachar}))
is indeed background gauge invariant.

As is well known, the deviation $h_{ab}$ transforms as 
\begin{equation}
 \delta_{\bar{\zeta}} h_{ab} = \bar{\nabla}_{a} \bar{\zeta}_{b} + \bar{\nabla}_{b} \bar{\zeta}_{a} \label{htra}
\end{equation}
under an infinitesimal diffeomorphism generated by a vector $\bar{\zeta}^{a}$. To see that $Q^{a}(\bar{\xi})$ is
background gauge invariant, we apply the gauge transformation (\ref{htra}) to $\Phi^{ab}_{L}$, the
invariance of which implies the background gauge invariance of ${\mathcal F}^{ab}$ (\ref{fab}) and hence of 
$Q^{a}(\bar{\xi})$. This is obviously easier since one can work with the linearized field equations $\Phi^{ab}_{L}$
rather than the gravitational charge $Q^{a}(\bar{\xi})$ which was not easy to find at the first place.

Some of the most basic expressions that are needed are as follows:
\begin{eqnarray*}
 \delta_{\bar{\zeta}} (\Gamma^{a}\,_{bc})_{L} & = & \bar{\zeta}^{e} \bar{R}_{e c}\,^{a}\,_{b} 
 + \bar{\nabla}_{c} \bar{\nabla}_{b} \bar{\zeta}^{a}, \\
 \delta_{\bar{\zeta}} (R^{a}\,_{bcd})_{L} & = & - \bar{R}^{a}\,_{bde} \bar{\nabla}_{c} \bar{\zeta}^{e} 
 -  \bar{\zeta}^{e} \bar{\nabla}_{c} \bar{R}^{a}\,_{bde} + \bar{R}^{a}\,_{bce} \bar{\nabla}_{d} \bar{\zeta}^{e}
 + \bar{\zeta}^{e} \bar{\nabla}_{d} \bar{R}^{a}\,_{bce} + \bar{R}_{b}\,^{e}\,_{cd} \bar{\nabla}_{e} \bar{\zeta}^{a} 
 + \bar{R}^{ae}\,_{cd} \bar{\nabla}_{b} \bar{\zeta}_{e}, \\
 \delta_{\bar{\zeta}} R_{ab}^{L} & = & \bar{\zeta}^{c} \bar{\nabla}_{c} \bar{R}_{ab} + \bar{R}_{ac} \bar{\nabla}_{b} \bar{\zeta}^{c} 
 + \bar{R}_{bc} \bar{\nabla}_{a} \bar{\zeta}^{c}, \\
 \delta_{\bar{\zeta}} R_{L} & = & \bar{\zeta}^{a} \bar{\nabla}_{a} \bar{R}.
\end{eqnarray*}
These easily follow considering the action of (\ref{htra}) on the expressions (\ref{chilin}) to (\ref{ricscalin}). One can similarly
calculate the analogous expressions regarding the action of (\ref{htra}) on (\ref{var}) (and on those ones that have not been listed).
Once again we have employed the Cadabra software \cite{cad1,cad2} at this stage and found that
\[ \delta_{\bar{\zeta}} \Phi_{ab}^{L} = \bar{\zeta}^{c} \bar{\nabla}_{c} \bar{\Phi}_{ab} 
 + \bar{\Phi}_{ac} \bar{\nabla}_{b} \bar{\zeta}^{c} + \bar{\Phi}_{bc} \bar{\nabla}_{a} \bar{\zeta}^{c}, \]
which vanishes thanks to \(\bar{\Phi}_{ab}(\bar{g}) = 0 \). As explained by the reasoning above, this shows that 
the gravitational charge $Q^{a}(\bar{\xi})$ is indeed background gauge invariant.

\section{\label{appc} Useful identities}
Throughout the calculation we have used a number of identities involving the Killing vectors. The proofs of these
are very simple, so we simply state them:
\[ \bar{\nabla}_{a} \bar{\nabla}_{b} \bar{\xi}_{c}  =  \bar{R}_{cbad} \bar{\xi}^{d}, \quad
 \bar{\nabla}^{b} \bar{\nabla}_{a} \bar{\xi}_{b} = \bar{R}_{ab} \bar{\xi}^{b}, \quad 
 \bar{\xi}^{a} \bar{\nabla}_{a} \bar{R} =  0, \quad
 \bar{\xi}^{c} \bar{\nabla}_{c} \bar{R}_{ab} = - ( \bar{R}_{ca} \bar{\nabla}_{b} +  \bar{R}_{cb} \bar{\nabla}_{a} ) \bar{\xi}^{c}. \]
The following are also useful identities involving the Riemann tensor. We state them without proof since they
follow from elementary properties of the Riemann tensor:
\[ \bar{\nabla}_{a} \bar{R}^{a}\,_{bcd} = \bar{\nabla}_{c} \bar{R}_{bd} - \bar{\nabla}_{d} \bar{R}_{bc}, \quad
 \bar{R}^{acde} \bar{R}_{bcde} = 2 \bar{R}^{acde} \bar{R}_{bdce}, \quad  
 \bar{R}^{abef} \bar{R}_{cfed} = \bar{R}^{afeb} \bar{R}_{cdef}. \]

\end{document}